# Quantum synapse for cold atoms[1]

G.A. Kouzaev and K.J Sand
Department of Electronics and Telecommunications
Norwegian University of Science and Technology-NTNU
7031 Trondheim, Norway
e-mail: guennadi.kouzaev@iet.ntnu.no

**Abstract.** In this paper, the quantum synaptic effect is studied that arisen in the system of two crossed wires excited by the static (DC) and radio-frequency (RF) currents. The potential barrier between the two orthogonal atom streams is controlled electronically and the atoms can be transferred from one wire to another under certain critical values of the RF and DC currents. The results are interesting in the study of quantum interferometry and quantum registering of cold atoms.

**Introduction.** Contemporary hopes to realize quantum computing are with the manipulation of cold atoms by DC, RF, and optical fields. The cold matter is the atoms cooled with the laser light and magnetic field to extremely low temperatures close to the absolute zero. Such atoms are interacting with each other and with the outer world according the rules dictated by quantum mechanics. This matter state is convenient to study the quantum mechanical laws and to realize the new devices like high-sensitive quantum interferometers and even registers for quantum computing.

In [1], an idea of cooling and control of the atom matter by strong static and magnetic fields was proposed. Such trapping is described by the dressed atom formalism and semi-analytical formulas [2,3]. A convenient form of this formalism was proposed in [4] where the dependence of the effective trapping potential on the direction of the RF field regarding to the static magnetic field was taken into account. Several traps have been studied by this approach, including the Ioffe-Pritchard trap excited by combined strong RF and static magnetic fields [4-7]. It was shown that this excitation allows realizing new spatial forms of the trapping potential.

For instance, the studied in [6,7] Ioffe-Pritchard trap confines two sorts of atoms at the difference to the DC excitation. The weak-field-seeking atoms concentrate at the effective potential minimums. The strong-field-seeking atoms stay at the trapping potential maximums. This trap excitation allows studying the interaction of atoms of different quantum states and the arisen quantum entanglement.

**Modeling Results.** This paper is on cold matter handling by the combined RF and static magnetic fields. It was studied the spatial shapes of the effective potential excited by the crossed wires carrying the RF and static currents.

Each wire is covered by atom clouds placed at the cylindrical surface where the combined effective potential is minimal. The atoms can be moved by the biasing static magnetic field. Normally, these two atom streams are isolated from each other by the potential barrier. The decreasing of the RF field allows increasing the radius of the cylindrical clouds, and they can touch each other. In this case, the atoms can wander from one wire to another. This effect is similar to the synapse effect found in the nervous system. There, the molecules are transferred through the synaptic slot under the control of the electrochemical potential. Some authors suppose that the quantum synaptic effect takes place even in the living matter [8], but the mechanism providing the long range quantum coherency at such high temperatures is still unknown.

Fig. 1 shows the studied geometry – two crossed wires carrying the RF and DC currents. A constant biasing magnetic field is added to avoid the Majorana flip-flops of atoms and realize pumping of the atom clouds.

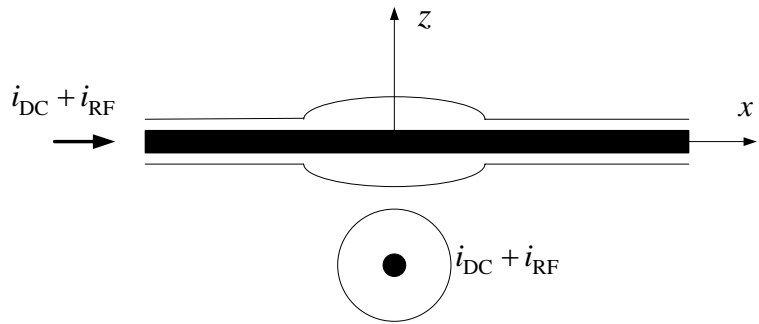

Fig. 1. Two crossed wires carrying the DC and RF currents and the spatial form of the minimum potential in general case.

---

The surface of the maximal concentration of cold atoms is deformed near the cross-point of the wires, but the atom clouds do not touch each other due to the potential barrier. Fig. 2 shows the spatial shape of the minimum surface under the critical condition. In this case, the potential barrier is zero , and the cold atoms cam be pumped from one wire to another.

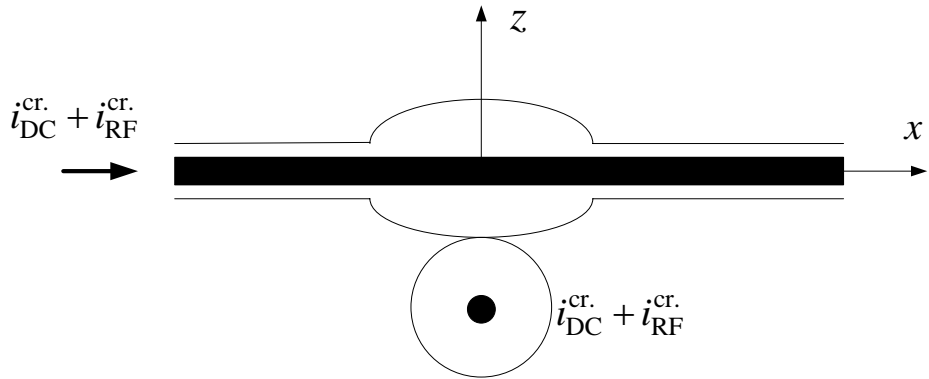

Fig. 2. Two crossed wires carrying the critical DC and RF currents and the spatial form of the minimum potential.

Fig. 3 shows a 3D view of the potential for the second case when the potential barrier is close to the zero. The effective potential function is calculated analytically in the assumption that the frequency is low ($f \leq 1$ MHz) and the diffraction effect is negligible at the point where two minimum potential surfaces are touching each other [6,7].

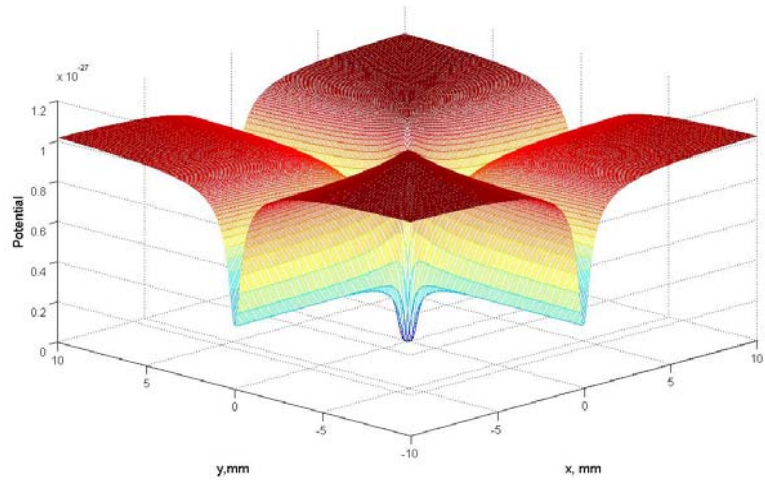

Fig. 3. The potential shape at the touching point of the minimum surfaces. Here, $f = 0.8$ MHz, $i_{RF} = 0.05$ A, $i_{DC} = 0.0925$ A.

**Conclusions.** A new effect was found in the system of two crossed wires carrying strong static and RF currents. The potential barrier in this 3D system can be controlled electronically, and the condition to transfer the cold atoms from one wire to another was found. The found effect is interesting in quantum interferometers and registers of cold atoms.